\documentstyle[12pt]{article}
\begin{document}
\begin{flushright}
arXiv: hep-th/0702039  \\ CAS-PHYS-BHU/Preprint
\end{flushright}

\vskip 2cm
\begin{center}
{\bf \large ABELIAN 2-FORM GAUGE THEORY: SUPERFIELD FORMALISM}

\vskip 2cm

{\bf R. P. Malik}\footnote{Presently also associated with the DST-Centre for
Interdisciplinary Mathematical Sciences, Faculty of Science, Banaras Hindu University,
Varanasi - 221 005 (U. P.), India.}\\
{\it Physics Department, Centre of Advanced Studies,}\\
 {\it Banaras Hindu University, Varanasi-221 005, U. P., India}\\
{\small E-mails: malik@bhu.ac.in ; rudra.prakash@hotmail.com}

\end{center}

\vskip 2cm

\noindent
{\bf Abstract:}
 We derive the off-shell nilpotent
Becchi-Rouet-Stora-Tyutin (BRST) and anti-BRST symmetry
transformations for {\it all} the fields of a free Abelian 2-form
gauge theory by exploiting the geometrical superfield approach to
BRST formalism. The above four (3 + 1)-dimensional (4D) theory is
considered on a (4, 2)-dimensional supermanifold parameterized by
the four even spacetime variables $x^\mu$ (with $\mu = 0, 1, 2,
3)$ and a pair of odd Grassmannian variables $\theta$ and
$\bar\theta$ (with $\theta^2 = \bar\theta^2 = 0, \theta \bar\theta
+ \bar\theta \theta = 0$). 
One of the salient features of our present investigation is that the above
nilpotent (anti-)BRST symmetry transformations turn out to be {\it absolutely}
anticommuting due to the presence of a Curci-Ferrari (CF) type of 
restriction. The latter condition emerges due to the application
of our present superfield formalism. The actual CF condition, as is well-known,
is the hallmark of a 4D non-Abelian 1-form gauge theory. 
We demonstrate that our present 4D Abelian 2-form gauge theory imbibes
some of the key signatures of the 4D non-Abelian 1-form gauge theory.
We briefly comment on the generalization of our supperfield approach
to the case of Abelian 3-form gauge theory in four (3 + 1)-dimensions
of spacetime.

\baselineskip=16pt

\vskip 1cm


\newpage

\noindent
\section{Introduction}

\noindent One of the most attractive and geometrically intuitive
theoretical approaches, that provides a glimpse of the
``physical'' understanding of the mathematical properties
associated with the nilpotent (anti-)BRST symmetries and their
corresponding generators (i.e. conserved and nilpotent charges),
is the superfield approach to BRST formalism (see,
e.g., [1-8]). 

In particular, the superfield approaches, proposed in
[3-6], are such that the geometrical interpretations for
(i) the nilpotent ($s_{(a)b}^2 = 0$) (anti-)BRST symmetry
transformations $s_{(a)b}$ (and their
corresponding nilpotent ($Q_{(a)b}^2 = 0$) and conserved
generators $Q_{(a)b}$), (ii) the nilpotency property ($s_{(a)b}^2
= 0, Q_{(a)b}^2 = 0$) itself, and (iii) the anticommutativity
property ($s_b s_{ab} + s_{ab} s_b = 0, Q_b Q_{ab} + Q_{ab} Q_b =
0$), etc., become very transparent\footnote{
We have chosen here the standard notations used in [9-24].}. These
results are the indispensable consequences of the superfield
formulation developed in [3-6].

The above 
superfield approaches (especially the ones in [3-6]) 
have been exploited in the context
of the gravitational theory and the 
(non-)Abelian 1-form ($A^{(1)} = dx^\mu A_\mu$) gauge
theories where the (anti-)BRST symmetry transformations for the
gauge and (anti-)ghost fields of the above theories 
have been derived very accurately. The
geometrical origin and interpretations for the nilpotent transformations 
(and their corresponding generators) have
also been provided within the 
framework of the above superfield formulations.

The key role, in the application of the above approaches [1-8] to 1-form gauge theories,
is played by the so-called horizontality condition where the
super curvature 2-form (i.e. $\tilde F^{(2)} = \tilde d \tilde
A^{(1)} + g \tilde A^{(1)} \wedge \tilde A^{(1)}$) is equated to
the ordinary curvature 2-form (i.e. $ F^{(2)} = d A^{(1)} + g
A^{(1)} \wedge A^{(1)}$). In the above, the symbol $\tilde d =
dx^\mu \partial_\mu + d \theta
\partial_\theta + d \bar\theta \partial_{\bar\theta}$ (with
$\tilde d^2 = 0$) is the super exterior derivative and $\tilde
A^{(1)}$ stands for the super 1-form connection defined on the (4,
2)-dimensional supermanifold that is parametrized by the four
spacetime variables $x^\mu$ (with $\mu = 0, 1, 2, 3$) and a pair
of Grassmannian variables $\theta$ and $\bar\theta$ (with
$\theta^2 = \bar\theta^2 = 0, \theta \bar\theta + \bar\theta
\theta = 0$). On the ordinary four (3 + 1)-dimensional (4D)
spacetime manifold (parametrized by the ordinary spacetime
variable $x^\mu$ alone), the ordinary exterior derivative $d =
dx^\mu \partial_\mu$ (with $d^2 = 0$) and the 1-form connection 
$A^{(1)} = dx^\mu A_\mu$ define the ordinary 2-form $F^{(2)}$.

In the expressions for the above
(super) 2-forms, $g$ is the coupling constant whose limiting case
(i.e. $g \to 0$) produces the horizontality condition for the 4D
Abelian 1-form gauge theory. This horizontality condition has been referred
to as the soul-flatness condition in [13] which amounts to setting
equal to zero all the Grassmannian components of the
(anti)symmetric curvature tensor that constitutes the super 2-form
$\tilde F^{(2)}$.

Recently, in a set of papers [14-24], the above superfield
approaches [3-6] have been consistently extended so as to derive
the nilpotent (anti-)BRST symmetry transformations that exist 
for the matter
fields together with such a set of nilpotent symmetry
transformations for the gauge and (anti-)ghost fields. The latter
set of transformations, as pointed out earlier, are derived due to
the application of the horizontality condition alone. We have
christened the extended version [14-24] of the above superfield
approaches [3-6] as the augmented superfield formalism. 
In this approach, in addition to the horizontality
condition (that is applied on the gauge superfield), a few
restrictions have been imposed on the matter as well as the gauge
superfields of the supersymmetric gauge theory [14-24].

In our very
recent works [21-24], we have been able to generalize the
horizontality condition itself where a {\it single} restriction,
on the superfields of the suitably chosen supermanifold, produces
{\it all} the nilpotent (anti-)BRST symmetry transformations for
{\it all} the fields of a given interacting (non-)Abelian 1-form
gauge theory without spoiling the cute geometrical interpretations
that emerge from the horizontality condition
alone.

It would be very nice endeavour to study the impact of the
geometrical superfield approach [3-6,14-24] in the context of the
{\it (non-)Abelian} 2-form gauge theories that have
become very popular and pertinent in the realm of modern
developments in the (super)string theories, related extended
objects and supergravity theories (see, e.g. [25-27]). As
a first modest step, we apply, in our present endeavour, the
geometrical superfield formulation to the 4D free {\it Abelian} 2-form
gauge theory and derive the off-shell 
nilpotent (anti-)BRST symmetry transformations for all the fields
of the theory. In addition, we provide their geometrical
interpretations in the language of the translational generators
along the Grassmannian directions of the appropriately chosen
supermanifolds.

There appear some novel features in the realm of
the application of the above superfield approach to the Abelian 2-form gauge
theory which do not crop up in the application of the very same
approach to the 4D Abelian 1-form gauge theory. For instance, we
obtain a CF
type of restriction on the 4D {\it bosonic} local fields of the theory
which enables us to obtain an absolutely anticommuting set of
(anti-)BRST symmetry transformations. 
It is to be noted that this type of restriction 
happens to be a key signature of the non-Abelian 1-form gauge theory 
where the bosonic and fermionic (ghost) fields participate in
the explicit form of the CF condition that ensures anticommutativity 
of the (anti-)BRST transformations [28].

The 4D free Abelian 2-form gauge theory, with its antisymmetric
(i.e. $B_{\mu\nu} = - B_{\nu\mu})$ gauge potential $B_{\mu\nu}$,
is interesting in its own right as it provides a dual description
of the massless scalar fields [29,30]; appears in the supergravity
multiplets [27] and excited states of the quantized (super)strings
[25,26]; plays a crucial role in the existence of the
noncommutative structure for string theory [31]; provides mass to
the 4D Abelian 1-form ($A^{(1)} = dx^\mu A_\mu$) gauge field
$A_\mu$ through a topological coupling (i.e. the celebrated $B
\wedge F$ term) where the $U(1)$ gauge invariance and mass
co-exist {\it together} without taking any recourse to the
presence of the Higgs fields, etc.

Furthermore, in our earlier 
works [32-34], we
have been able to show that the 4D Abelian 2-form gauge theory
provides

(i) an interesting field theoretical model for the Hodge
theory [32,33] because all the de Rham cohomological operators
find their analogue(s) in the language of the conserved charges
and the continuous symmetry transformations they generate,

(ii) a tractable model where the connection between the
gauge symmetry and the translation subgroup of
the Wigner's little group turns out to be quite transparent [34], and

(iii) a gauge field theoretic model for the quasi-topological
field theory [34].

Thus, it is important to know about this gauge
potential and the corresponding gauge theory from various points
of view. Our present endeavour is an attempt in that direction.

The purpose of the present paper is to study the geometrical
structure behind the nilpotent (anti-)BRST symmetry
transformations (and their corresponding generators) that are
associated with the 4D free Abelian 2-form gauge theory in the
framework of the superfield approach to BRST formalism. 
We exploit the power of
the gauge (i.e. (anti-)BRST) invariant horizontality condition
 to derive the off-shell nilpotent
(anti-)BRST symmetry transformations for all the {\it basic} fields of the
appropriate Lagrangian densities (cf. (2.4), (2.5) below). The nilpotent
transformations for the {\it auxiliary} fields are determined by the
requirement of the absolute anticommutativity ($s_b s_{ab} + s_{ab} s_b = 0$)
of the (anti-)BRST symmetry transformations $s_{(a)b}$ (when they
act on any field of the theory).

One of the key 
results of our present investigation is the derivation of the CF
type restriction (cf. (3.12) below)  within the framework of the
superfield approach to BRST formalism. In fact, it is because of our present
investigation that we were able to derive an absolutely anticommuting set of
(anti-)BRST symmetry transformations in the case of 4D free Abelian
2-form gauge theory [35]. In this work, we were also able to demonstrate that the analogue
of the CF restriction (cf. (2.8)) would always be required for the 
derivation of the above kind of anticommuting symmetry transformations 
in the context of higher p-form ($p \geq 2$) Abelian gauge theories. It was
also claimed that there was a deep connection between the restriction (2.8)
and the concept of gerbes [35].

Our present investigation is interesting and essential on the
following grounds.

First and foremost, to the best of our
knowledge, the geometrical superfield approach to BRST formalism (especially 
proposed in [3-6,14-24]) has never been applied to the 2-form (and/or higher
form) gauge theories.

Second, one of the
key features of our present superfield approach is the derivation of the
nilpotent (anti-)BRST symmetry transformations that {\it always} turn
out to be absolutely anticommuting\footnote{The absolute anticommutativity
property encodes the linear independence of the nilpotent (anti-)BRST symmetry transformations
which emerge from a given ``classical'' local gauge symmetry transformation.}. 
As a result, it is important for us to
apply the superfield formulation to the 4D Abelian 2-form gauge theory
where the known (anti-)BRST symmetry transformations were {\it not}
absolutely anticommuting in nature [10,32,33] (see, e.g. subsection 2.1 below).

Third, it is for the
first time, that we are coming across a CF type of restriction in the context
of an {\it Abelian} gauge theory for the proof of the anticommutativity
of the (anti-)BRST transformations. The derivation of this restriction 
is a completely new result.

Finally, our present
endeavour is our first modest step towards our main goal of 
applying the superfield approach to the 4D non-Abelian 2-form gauge theory,
higher p-form ($p \geq 3$) gauge theories as well as the gravitational
theories.

The contents of our present paper are organized as follows.

In
section 2, we discuss the bare essentials of the off-shell nilpotent and
(i) anticommuting up to a U(1) vector gauge transformation, as well as
(ii) absolutely anticommuting (anti-)BRST symmetry transformations 
for the 4D free Abelian 2-form gauge theory in the framework of Lagrangian 
formulation to set up the notations and conventions.

The latter 
(anti-)BRST symmetry transformations and the CF type restriction 
(cf. (2.8) below) are derived in section 3 by exploiting a
gauge-invariant restriction on the super 2-form gauge connection that
are defined on the (4, 2)-dimensional supermanifold.

Section 4 deals with
the (anti-)BRST invariance of the appropriate Lagrangian
densities of the present theory in the language of the superfield formalism.

Finally, in
section 5, we summarize our key results, make some concluding
remarks and point out a few future directions for further investigations.

Our Appendix A deals concisely with the generalization of our superfield approach
to the Abelian 3-form gauge theory in four (3 + 1)-dimension of
spacetime.

\section{Preliminary: off-shell nilpotent (anti-)BRST symmetry transformations in Lagrangian
formulation}

\noindent
Here we discuss briefly the off-shell nilpotent (anti-)BRST symmetry transformations
for the 4D free Abelian 2-form gauge theory where 
(i) the transformations are  
anticommuting up to a U(1) vector gauge transformation, and 
(ii) the above transformations are absolutely anticommuting due to a
specific restriction on the fields of the theory. 

\subsection{Non-anticommuting but off-shell nilpotent (anti-)BRST symmetry transformations}

\noindent We begin with the following off-shell nilpotent
(anti-)BRST invariant Lagrangian density of the 4D\footnote{We adopt here the 
conventions such that the 4D flat
Minkowskian metric is a diagonal metric with the signatures
 $(+1, -1, -1, -1)$. This implies that $(P
\cdot Q) = P^\mu Q_\mu \equiv \eta_{\mu\nu} P^\mu Q^\nu = P_0 Q_0
- P_i Q_i$ corresponds to the dot product between two non-null four-vectors
$P_\mu$ and $Q_\mu$ where the Greek indices $\mu, \nu, \kappa....=
0, 1, 2, 3$ stand for the spacetime directions on the 4D
Minkowskian spacetime manifold and Latin indices $i, j, k....= 1,
2, 3$ denote only the space directions on the above spacetime
manifold.}
free Abelian 2-form gauge theory (see, e.g., [10,32,33]): $$
\begin{array}{lcl}
{\cal L}_B &=& {\displaystyle \frac{1}{12}\; H^{\mu\nu\kappa}
H_{\mu\nu\kappa} + B^\mu \Bigl (\partial^\nu B_{\nu\mu}
-
\partial_\mu \phi \Bigr ) - \frac{1}{2} B^\mu B_\mu
- \partial_\mu \bar \beta\;
\partial^\mu \;\beta} \nonumber\\ &+&
{\displaystyle \Bigl (\partial_\mu \bar C_\nu -
\partial_\nu \bar C_\mu \Bigr )\;
(\partial^\mu C^\nu) + \rho \;\Bigl (\partial \cdot C + \lambda
\Bigr ) + \Bigl (\partial \cdot \bar C + \rho \Bigr )\; \lambda },
\end{array} \eqno(2.1)
$$ where the totally antisymmetric field strength tensor
$H_{\mu\nu\kappa} = \partial_\mu B_{\nu\kappa} + \partial_\nu
B_{\kappa\mu} + \partial_\kappa B_{\mu\nu}$ is derived from the
3-form $H^{(3)} = d B^{(2)} \equiv (1/3!) (dx^\mu \wedge dx^\nu
\wedge dx^\kappa) H_{\mu\nu\kappa}$ that is constructed with the
help of the nilpotent ($d^2 = 0$) exterior derivative $d = dx^\mu
\partial_\mu$ and the 2-form connection $B^{(2)} = (1/2!)
(dx^\mu \wedge dx^\nu) B_{\mu\nu}$. The latter defines the
antisymmetric potential (i.e. the gauge field) $B_{\mu\nu}$ of the
present 4D free Abelian 2-form gauge theory. The bosonic auxiliary
field $B_\mu = + (\partial^\nu B_{\nu\mu} -
\partial_\mu \phi)$ is the Nakanishi-Lautrup auxiliary field.

It will be noted that
there are a pair of fermionic (i.e. $\rho^2 = 0, \lambda^2 = 0,
\rho \lambda + \lambda \rho = 0$) Lorentz scalar auxiliary ghost
fields $\rho = - \frac{1}{2} (\partial \cdot \bar C) $ and
$\lambda = - \frac{1}{2} (\partial \cdot C)$ in the theory.
Furthermore, there exists a set of fermionic ($C_\mu^2 = 0,
\bar C_\mu^2 = 0, C_\mu \bar C_\nu + \bar C_\nu C_\mu = 0,$ etc.)
Lorentz vector (anti-)ghost fields $(\bar C_\mu) C_\mu$ (with the
ghost number $\mp 1$) and a pair of bosonic (i.e. $\beta^2 \neq 0,
\bar\beta^2 \neq 0, \beta \bar \beta = \bar\beta \beta$) Lorentz
scalar (anti-)ghost fields $(\bar\beta) \beta$ (with the ghost
number $\mp 2$) in the theory. These (anti-)ghost fields are
required for the gauge-fixing term (i.e. $+ (1/2) (\partial^\nu
B_{\mu\nu} - \partial_\mu \phi)^2$) that is present in the
Lagrangian density (2.1) of the present theory. The field
$\phi$, that appears in the gauge-fixing term, is a massless
$\Box \phi = 0$ scalar field where $\Box =
\partial_0^2 - \partial_i^2$. This field is required due to
the stage-one reducibility in the theory.

The above Lagrangian density for the free Abelian 2-form gauge
theory is endowed with the following off-shell nilpotent
($\tilde s_{(a)b}^2 = 0$) but {\it not} absolutely anticommuting
($\tilde s_b \tilde s_{ab} + \tilde s_{ab} \tilde s_b \neq 0$)
(anti-)BRST symmetry transformations $\tilde s_{(a)b}$, namely;
$$
\begin{array}{lcl}
\tilde s_b B_{\mu\nu} &=& - (\partial_\mu C_\nu - \partial_\nu C_\mu),
\qquad \tilde s_b C_\mu = - \partial_\mu \beta, \qquad \tilde s_b \bar C_\mu =
\;-\;B_\mu, \nonumber\\ \tilde s_b \phi &=&  \lambda, \qquad \tilde s_b \bar
\beta = - \rho, \qquad \tilde s_b \Bigl [\rho, \lambda, \beta,  B_\mu,
H_{\mu\nu\kappa} \Bigr ] = 0,
\end{array} \eqno(2.2)
$$ $$
\begin{array}{lcl}
\tilde s_{ab} B_{\mu\nu} &=& - (\partial_\mu \bar C_\nu - \partial_\nu
\bar C_\mu), \qquad \tilde s_{ab} \bar C_\mu = + \partial_\mu \bar
\beta, \qquad \tilde s_{ab}  C_\mu = \;+\; B_\mu, \nonumber\\ \tilde s_{ab}
\phi &=&
 \rho, \qquad \tilde s_{ab}  \beta = - \lambda, \qquad \tilde s_{ab} \Bigl [\rho,
\lambda, \bar \beta, B_\mu, H_{\mu\nu\kappa} \Bigr ] = 0.
\end{array} \eqno(2.3)
$$ It will be noted that:

(i) The above nilpotent (anti-)BRST symmetry
transformations differ from the ones, given
in our earlier works [32,33], by a sign factor. The above choice
has been taken only for the algebraic convenience.

 (ii) Under the above nilpotent symmetry
transformations, the Lagrangian density transforms as: $\tilde s_b
{\cal L}_B = - \partial_\mu [ B^\mu \lambda + (\partial^\mu C^\nu
- \partial^\nu C^\mu) B_\nu - \rho
\partial^\mu \beta]$ and $\tilde s_{ab} {\cal L}_B = - \partial_\mu [
B^\mu \rho + (\partial^\mu \bar C^\nu - \partial^\nu \bar C^\mu)
B_\nu - \lambda
\partial^\mu \bar\beta]$. Thus, ${\cal L}_B$ is quasi-invariant under
(2.2) and (2.3).

(iii) The anticommutativity property
$\tilde s_b \tilde s_{ab} + \tilde s_{ab} \tilde s_b = 0$ is 
{\it not} precisely valid for
the (anti-)ghost fields $(\bar C_\mu)C_\mu$ because 
$ \tilde s_b \tilde s_{ab}
\bar C_\mu = - \partial_\mu \rho$ but $\tilde s_{ab} \tilde s_b \bar C_\mu = 0$.
Furthermore, we have 
$\tilde s_b \tilde s_{ab} C_\mu = 0$ but
$\tilde s_{ab} \tilde s_b C_\mu = + \partial_\mu
\lambda$. Thus, the above (anti-)BRST transformations $\tilde s_{(a)b}$ 
in (2.2) and (2.3) are {\it not} absolutely anticommuting in nature.

 (iv) The above anticommutativity property is valid up
to an Abelian vector gauge transformation [i.e. $(\tilde s_b \tilde s_{ab} 
+ \tilde s_{ab} \tilde s_b) \bar C_\mu = - \partial_\mu \rho, 
(\tilde s_b \tilde s_{ab} + \tilde s_{ab} \tilde s_b) C_\mu = +
\partial_\mu \lambda$] for the fermionic vector (anti-)ghost fields $(\bar
C_\mu)C_\mu$ because they transform up to a total derivative term.

(v) The above observation is totally different
from the anticommutativity property that is found for the 4D
(non-)Abelian 1-form gauge theories. To be precise, the
anticommutativity property is very much sacrosanct in the case of
the 4D (non-)Abelian 1-form gauge theories where the (anti-)ghost
fields are {\it only} fermionic in nature and they are found to be
Lorentz scalars {\it only}. The vector (anti-)ghost fields do not
exist in these theories.

(vi) The absolute anticommutatvity, in the case of the non-Abelian
1-form gauge theory, emerges only due to the presence of the Curci-Ferrari
restriction [28]. The superfield formulation, developed in [3], leads
to the explicit derivation of the above restriction.

(vii) The superfield approaches [3-6, 14-24] can never be able
to capture the nilpotent symmetry transformations (2.2) and (2.3)
because the latter are not absolutely anticommuting in nature.
The absolute anticommutatvity, however, is a key consequence of [3-6,14-24].

\subsection{Absolutely anticommuting and nilpotent (anti-)BRST symmetry transformations}

\noindent
Let us begin with the modified versions (${\cal L}_B^{(b)}, {\cal L}_B^{(ab)}$)
of the Lagrangian density
(2.1) for the free 4D Abelian 2-form gauge theory [35] 
$$
\begin{array}{lcl}
{\cal L}^{(b)}_B &=& {\displaystyle \frac{1}{12}\; H^{\mu\nu\kappa}
H_{\mu\nu\kappa} + B^\mu \Bigl (\partial^\nu B_{\nu\mu}) 
+ \frac{1}{2} \Bigl (B \cdot B + \bar B \cdot \bar B \Bigr )
+ \partial_\mu \bar \beta\;
\partial^\mu \;\beta} \nonumber\\ &+&
{\displaystyle \Bigl (\partial_\mu \bar C_\nu -
\partial_\nu \bar C_\mu \Bigr )\;
(\partial^\mu C^\nu) + \Bigl (\partial \cdot C - \lambda
\Bigr )\; \rho + \Bigl (\partial \cdot \bar C + \rho \Bigr )\; \lambda }
\nonumber\\
&+& L^\mu \Bigl (B_\mu - \bar B_\mu - \partial_\mu \phi)
- {\displaystyle \frac{1}{2}} \partial^\mu \phi \partial_\mu \phi,
\end{array} \eqno(2.4)
$$ 
$$
\begin{array}{lcl}
{\cal L}^{(ab)}_B &=& {\displaystyle \frac{1}{12}\; H^{\mu\nu\kappa}
H_{\mu\nu\kappa} + \bar B^\mu \Bigl (\partial^\nu B_{\nu\mu}) 
+ \frac{1}{2} \Bigl (B \cdot B + \bar B \cdot \bar B \Bigr )
+ \partial_\mu \bar \beta\;
\partial^\mu \;\beta} \nonumber\\ &+&
{\displaystyle \Bigl (\partial_\mu \bar C_\nu -
\partial_\nu \bar C_\mu \Bigr )\;
(\partial^\mu C^\nu) + \Bigl (\partial \cdot C - \lambda
\Bigr )\; \rho + \Bigl (\partial \cdot \bar C + \rho \Bigr )\; \lambda }
\nonumber\\
&+& L^\mu \Bigl (B_\mu - \bar B_\mu - \partial_\mu \phi)
- {\displaystyle \frac{1}{2}} \partial^\mu \phi \partial_\mu \phi,
\end{array} \eqno(2.5)
$$ 
where an additional auxiliary vector field $\bar B_\mu$ and the Lagrange
multiplier field $L_\mu$ have been introduced. The above Lagrangian
densities
respect the off-shell nilpotent and 
{\it absolutely} anticommuting 
($s_b s_{ab} + s_{ab} s_b = 0$) (anti-)BRST symmetry transformations
$s_{(a)b}$ 
$$
\begin{array}{lcl}
s_b B_{\mu\nu} &=& - (\partial_\mu C_\nu - \partial_\nu C_\mu),
\quad s_b C_\mu = - \partial_\mu \beta, \quad s_b \bar C_\mu =
\;-\;B_\mu,  \quad s_b L_\mu = - \partial_\mu \lambda,
\nonumber\\  s_b \phi &=&  \lambda, \qquad  s_b \bar
\beta = - \rho, \qquad s_b \bar B_\mu = - \partial_\mu \lambda,
\qquad s_b \Bigl [\rho, \lambda, \beta,  B_\mu,
H_{\mu\nu\kappa} \Bigr ] = 0,
\end{array} \eqno(2.6)
$$ $$
\begin{array}{lcl}
s_{ab} B_{\mu\nu} &=& - (\partial_\mu \bar C_\nu - \partial_\nu
\bar C_\mu), \quad s_{ab} \bar C_\mu = - \partial_\mu \bar
\beta, \quad s_{ab}  C_\mu = + \bar B_\mu, \quad s_{ab} L_\mu 
= - \partial_\mu \rho,  \nonumber\\ s_{ab} \phi &=&
 \rho, \qquad s_{ab}  \beta = - \lambda, \qquad 
s_{ab} B_\mu = \partial_\mu \rho, \qquad  s_{ab} \Bigl [\rho,
\lambda, \bar \beta, \bar B_\mu, H_{\mu\nu\kappa} \Bigr ] = 0.
\end{array} \eqno(2.7)
$$ To be more precise,
it can be checked that the above nilpotent 
transformations are absolutely anticommuting
(i.e. $\{ s_b, s_{ab} \} B_{\mu\nu} = 0$) if and only if the following
constrained surface\footnote{This relation actually owes its origin
to our present work (cf. equation (3.12) below).}, defined in terms of the 4D local fields,
is satisfied, namely;
$$
\begin{array}{lcl}
\bar B_\mu (x)  - B_\mu (x) + \partial_\mu \phi (x) = 0.
 \end{array}\eqno(2.8)
$$
It is elementary to check that, for the rest of the fields of the theory,
there is no need of the constrained equation (2.8) for the proof of the
anticommutativity $s_b s_{ab} + s_{ab} s_b = 0$.

The noteworthy points, at this stage, are

(i) The Lagrangian densities (2.4) and (2.5) are equivalent on the constrained
surface (2.8). This latter equation is the analogue of the Curci-Ferrari condition [28]
that is invoked in the 4D non-Abelian 1-form gauge theory for the proof of the
anticommutativity of the off-shell nilpotent (anti-)BRST symmetry transformations.

(ii) The constrained condition (2.8) could be derived from (2.4) and (2.5) as 
an equation of motion with respect to the multiplier field $L_\mu$. Furthermore,
it can be checked that the restriction (2.8) is an (anti-)BRST invariant quantity
(i.e. $s_{(a)b} [\bar B_\mu -  B_\mu + \partial_\mu \phi] = 0$).

(iii) Under the absolutely anticommuting (anti-)BRST symmetry transformations,
the Lagrangian densities (2.4) and (2.5) transform as:
$s_b {\cal L}^{(b)} = - \partial_\mu [(\partial^\mu C^\nu - \partial^\nu C^\mu)
B_\nu + \lambda B^\mu + \rho \partial^\mu \beta ],  
s_{ab} {\cal L}^{(ab)} = - \partial_\mu [(\partial^\mu \bar C^\nu 
- \partial^\nu \bar C^\mu)
\bar B_\nu - \rho \bar B^\mu + \lambda \partial^\mu \bar \beta ].  
$

(iv) Unlike the Curci-Ferrari condition of the 4D non-Abelian 1-form gauge theory [28]
where the auxiliary fields and the (anti-)ghost fields are  connected, the condition 
(2.8) invokes a relationship where the auxiliary fields and the derivative on the scalar field are taken into account. The latter relationship is deeply linked with the concept
of gerbes [35].

(v) The above off-shell nilpotent and absolutely anticommuting
(anti-)BRST symmetry transformations are generated by the conserved charges
$Q_{(a)b}$. For a generic
field $\Omega (x)$, this statement can be succinctly expressed in
the mathematical form, as $$
\begin{array}{lcl}
s_r \Omega (x) = - i\; [\Omega (x), Q_r]_{(\pm)},\; \;
\qquad \;r = b, ab,
\end{array}\eqno (2.9)
$$ where the $(\pm)$ signatures, as the subscripts on the above
square brackets, correspond to the (anti)commutators for the
generic field $\Omega (x)$, of the Lagrangian densities (2.4)
and/or (2.5), being fermionic(bosonic) in nature.

\section{Off-shell nilpotent and absolutely anticommuting
(anti-)BRST symmetries: superfield approach}

\noindent To derive the off-shell nilpotent (anti-)BRST symmetry
transformations (cf. (2.6),(2.7)) and the Curci-Ferrari type restriction\footnote{
It will be noted that the off-shell nilpotent and
non-anticommuting transformations (2.2) and (2.3) cannot
be derived by exploiting our present superfield formulation [3-6,14-24]. The
absolute anticommutativity and nilpotency of the (anti-)BRST symmetry transformations
are the key consequences of our geometrical superfield approach to BRST formalism.}
(2.8), we begin with the
superfields (that are the generalization of the basic 4D local
fields of the Lagrangian densities (2.4) and (2.5)) on an appropriately chosen 
(4, 2)-dimensional supermanifold, parameterized by the superspace
variable $Z^M = (x^\mu, \theta, \bar\theta)$. These superfields
can be expanded along the Grassmannian directions of the above supermanifold 
in terms of the basic fields of the Lagrangian densities
(2.4)/(2.5) and some extra secondary fields as 
 $$
\begin{array}{lcl}
\tilde {\cal B}_{\mu\nu} (x, \theta, \bar\theta) &=& B_{\mu\nu}
(x) + \theta\; \bar R_{\mu\nu} (x) + \bar\theta\; R_{\mu\nu} (x) +
i \;\theta \; \bar\theta\; S_{\mu\nu} (x), \nonumber\\ \tilde \beta
(x, \theta, \bar\theta ) &=& \beta (x) + \theta \;\bar f_1 (x) +
\bar\theta\; f_1 (x) + i\; \theta\; \bar\theta\; b_1 (x),
\nonumber\\ \tilde {\bar \beta} (x, \theta, \bar\theta) &=& 
\bar\beta (x) + \theta \;\bar f_2 (x) + \bar \theta\; f_2 (x) + i
\;\theta\;\bar\theta\; b_2 (x), \nonumber\\ \tilde \Phi (x,
\theta, \bar\theta) &=& \phi (x) + \theta \;\bar f_3 (x) +
\bar\theta\; f_3 (x) + i \;\theta \;\bar\theta\; b_3 (x),
\nonumber\\ \tilde {\cal F}_\mu (x, \theta, \bar\theta) &=& C_\mu
(x) + \theta \;\bar B^{(1)}_\mu (x) + \bar\theta\; B^{(1)}_\mu (x)
+ i \;\theta \; \bar\theta\;f^{(1)}_\mu (x), \nonumber\\ \tilde
{\bar {\cal F}}_\mu (x, \theta, \bar\theta) &=& \bar C_\mu (x) +
\theta\; \bar B^{(2)}_\mu (x) + \bar\theta\; B^{(2)}_\mu (x) + i
\; \theta\; \bar\theta \bar f^{(2)}_\mu (x).
\end{array}\eqno (3.1)
$$ In the limit $(\theta,
\bar\theta) \to 0$, we retrieve the basic 4D fields of the
Lagrangian densities (2.4)/(2.5) and the number of the fermionic 
(e.g. $R_{\mu\nu}, \bar R_{\mu\nu}, f_1, \bar f_1, f_2, \bar f_2,
f_3, \bar f_3, C_\mu, \bar C_\mu, f_\mu^{(1)}, \bar f_\mu^{(2)}$) and
bosonic (e.g. $B_{\mu\nu},  S_{\mu\nu}, b_1, \beta, b_2, \bar \beta,
b_3, \phi, B^{(1)}_\mu, \bar B^{(1)}_\mu, B_\mu^{(2)}, \bar B_\mu^{(2)}$)
fields on the r.h.s. of the above expansion do match. Thus, the sanctity
of the supersymmetry is maintained.

We have to exploit now the mathematical potential of the
horizontality condition (i.e. $\tilde d \tilde B^{(2)} = d
B^{(2)}$)\footnote{The general form of super 2-form connection 
$\tilde B^{(2)}$, on the (4, 2)-dimensional
supermanifold,  is: $\tilde B^{(2)} = (1/2!)\; (d Z^M \wedge d
Z^N) \tilde B_{MN} \equiv (1/2!) (dx^\mu \wedge dx^\nu)\;
\tilde B_{\mu\nu} + (dx^\mu \wedge d\theta)\; \tilde
B_{\mu\theta} + (dx^\mu \wedge d \bar\theta)
\tilde B_{\mu\bar\theta} + (1/2!) (d\theta
\wedge d\theta)\; \tilde B_{\theta\theta} + (1/2!) (d
\bar\theta \wedge d\bar\theta) \;\tilde B_{\bar\theta\bar\theta} +
(d \theta \wedge d\bar\theta) \tilde B_{\theta\bar\theta}
$ where we have chosen $\tilde B_{\mu\nu} = \tilde {\cal
B}_{\mu\nu} (x, \theta,\bar\theta)$, $\tilde B_{\mu\theta} =
\tilde {\bar {\cal F}}_\mu (x, \theta, \bar\theta)$, $\tilde
B_{\mu\bar\theta} = \tilde {\cal F}_\mu (x,\theta,\bar\theta)$,
$\tilde B_{\theta\bar\theta} = \tilde \Phi
(x,\theta,\bar\theta)$, $(1/2!) \tilde B_{\theta\theta} = \tilde
{\bar \beta} (x,\theta,\bar\theta)$ and $(1/2!)
B_{\bar\theta\bar\theta} = \tilde \beta (x,\theta,\bar\theta)$.}
to obtain the off-shell nilpotent and absolutely anticommuting 
(anti-)BRST symmetry
transformations of (2.6) and (2.7). To this end in mind, we first
of all, generalize the ordinary exterior derivative $d = dx^\mu
\partial_\mu$ as well as the 2-form $B^{(2)} = (1/2!) (dx^\mu
\wedge dx^\nu) B_{\mu\nu}$ of the ordinary 4D spacetime manifold
to their counterparts on the (4, 2)-dimensional supermanifold.
These are $$
\begin{array}{lcl}
&&d \to \tilde d = d Z^M \partial_{Z^M} \equiv dx^\mu
\partial_\mu + d \theta
\partial_{\theta} + d \bar\theta \partial_{\bar\theta},\; \qquad
 \partial_\mu \to \partial_{Z^M} = (\partial_\mu,
\partial_\theta, \partial_{\bar\theta}), \nonumber\\ && B^{(2)}
\to \tilde B^{(2)} = {\displaystyle \frac{1}{2!}} (dx^\mu \wedge
dx^\nu) \; \tilde {\cal B}_{\mu\nu} (x, \theta, \bar\theta)
\nonumber\\ && +(dx^\mu \wedge d \theta)\; \tilde {\bar {\cal
F}}_\mu (x, \theta, \bar\theta) + (dx^\mu \wedge d \bar\theta)\;
\tilde {\cal F}_\mu (x, \theta, \bar\theta) \nonumber\\ && +
(d\theta \wedge d \theta)\; \tilde {\bar \beta} (x, \theta,
\bar\theta) + (d \bar\theta \wedge d \bar\theta) \;\tilde \beta
(x, \theta, \bar\theta) + (d \theta \wedge d\bar\theta)\; \tilde
\Phi  (x, \theta, \bar\theta).
\end{array}\eqno (3.2)
$$ Taking the help of (3.1) and (3.2), it can be readily seen that
the above definitions (on the (4, 2)-dimensional supermanifold)
reduce to their counterparts (i.e. $d, B^{(2)})$ on the ordinary
4D spacetime manifold in the limit $(\theta, \bar\theta) \to 0$.

The horizontality condition is a gauge invariant
restriction because $d B^{(2)} = H^{(3)} \equiv (1/3!) (dx^\mu
\wedge dx^\nu \wedge dx^\kappa) H_{\mu\nu\kappa}$ is an (anti-)BRST
invariant quantity in the sense that $s_{(a)b} H_{\mu\nu\kappa} = 0$.
To see the consequences of the horizontality condition in its full
blaze of glory, we have to compute explicitly the super 3-form
$\tilde d \tilde B^{(2)}$ and set all the Grassmannian components
equal to zero. To this end in mind, we have the following explicit
expression for $\tilde d \tilde B^{(2)}$:
 $$
\begin{array}{lcl}
\tilde d \tilde B^{(2)} &=& {\displaystyle \frac{1}{2!}}
(dx^\kappa \wedge d x^\mu \wedge dx^\nu) (\partial_\kappa \tilde
{\cal B}_{\mu\nu}) + (d \theta \wedge d \theta \wedge d \theta)
(\partial_{\theta} \tilde {\bar \beta}) + (d \bar\theta \wedge d
\bar \theta \wedge d \bar\theta) (\partial_{\bar\theta} \tilde
\beta) \nonumber\\ &+& (d\theta \wedge d\bar\theta \wedge
d\bar\theta) \;\bigl [\partial_{\bar\theta} \tilde \Phi +
\partial_\theta \tilde \beta \bigr ] +
(d\bar \theta \wedge d\theta \wedge d\theta)\; \bigl
[\partial_{\theta} \tilde \Phi +
\partial_{\bar\theta} \tilde {\bar\beta} \bigr ]\nonumber\\
&+& {\displaystyle \frac{1}{2!}} (dx^\mu \wedge dx^\nu \wedge d
\theta)\; \bigl [
\partial_{\theta} \tilde {\cal B}_{\mu\nu} + \partial_\mu
\tilde {\bar {\cal F}}_\nu -
\partial_\nu \tilde {\bar {\cal F}}_\mu \bigr ] \nonumber\\
&+& (dx^\mu \wedge d \theta \wedge d \theta) \;\bigl
[\partial_{\theta} \tilde {\bar {\cal F}}_\mu +
\partial_\mu \tilde {\bar \beta} \bigr ]
+ (dx^\mu \wedge d \bar\theta \wedge d \bar \theta) \;\bigl
[\partial_{\bar\theta} \tilde {\cal F}_\mu +
\partial_\mu \tilde  \beta \bigr ]\nonumber\\
&+& {\displaystyle \frac{1}{2!}} (dx^\mu \wedge dx^\nu \wedge d
\bar \theta)\; \bigl [
\partial_{\bar\theta} \tilde {\cal B}_{\mu\nu} + \partial_\mu
\tilde {\cal F}_\nu -
\partial_\nu \tilde  {\cal F}_\mu \bigr ] \nonumber\\
&+& (dx^\mu \wedge d\theta \wedge d\bar\theta)\; \bigl [
\partial_\mu \tilde \Phi + \partial_\theta \tilde {\cal F}_\mu +
\partial_{\bar\theta} \tilde {\bar {\cal F}}_\mu \bigr ].
\end{array}\eqno (3.3)
$$ The first term is the above expression has to be equated with
the r.h.s (i.e. $d B^{(2)}$). This equality, in its full bloom, is as follows
$$
\begin{array}{lcl}
&& {\displaystyle \frac{1}{3!}} (dx^\mu \wedge dx^\nu \wedge
dx^\kappa) \; \bigl [
\partial_\mu \tilde {\cal B}_{\nu\kappa} (x, \theta, \bar\theta)
+ \partial_\nu \tilde {\cal B}_{\kappa\mu} (x, \theta, \bar\theta)
+ \partial_\kappa \tilde {\cal B}_{\mu\nu} (x, \theta, \bar\theta)
\bigr ] = \nonumber\\ && {\displaystyle \frac{1}{3!}} (dx^\mu
\wedge dx^\nu \wedge dx^\kappa) \; \bigl [
\partial_\mu B_{\nu\kappa} (x)
+ \partial_\nu B_{\kappa\mu} (x) +
\partial_\kappa B_{\mu\nu} (x) \bigr ].
\end{array}\eqno (3.4)
$$ It is clear that the l.h.s. of the above equation would have
some coefficients of the Grassmannian variables $\theta$,
$\bar\theta$ and $\theta\bar\theta$. These ought to be
zero for the sanctity of the horizontality condition 
($\tilde d  \tilde B^{(2)} = d B^{(2)}$) because the
r.h.s. of (3.4) does not contain any such kind of terms. To demonstrate this,
we proceed, purposely step-by-step, so that all the key
points of our computation could become clear.

Let us, first of all, set the coefficients of the 3-form
differentials $(d\theta \wedge d\theta \wedge d\theta)$ and $(d
\bar\theta \wedge d\bar\theta \wedge d\bar\theta)$ equal to zero.
These restrictions imply the following$$
\begin{array}{lcl}
&&\partial_\theta \tilde {\bar \beta} (x, \theta, \bar\theta) = 0
\Rightarrow \bar f_2 (x) = 0,\; \quad b_2 (x) = 0, \nonumber\\
&&\partial_{\bar\theta} \tilde \beta (x, \theta, \bar\theta) = 0
\Rightarrow  f_1 (x) = 0,\; \quad b_1 (x) = 0,
\end{array}\eqno (3.5)
$$ which entail upon the above superfields to reduce to $$
\begin{array}{lcl}
\tilde \beta (x,\theta,\bar\theta) \to
\tilde \beta^{(r)} (x, \theta) = \beta (x) + \theta \bar f_1 (x),\;
\;\; \;\tilde {\bar \beta} (x,\theta,\bar\theta) \to 
\tilde {\bar \beta}^{(r)} (x, \bar\theta) =  \bar \beta
(x) + \bar \theta  f_2 (x).
\end{array}\eqno (3.6)
$$  We go a step further and set the
coefficients of the differentials $(d\theta \wedge d\bar\theta
\wedge d\bar\theta)$ and $(d \bar\theta \wedge d\theta \wedge d
\theta)$ equal to zero. This condition leads to the following
relationships$$
\begin{array}{lcl}
&&\partial_{\bar\theta} \tilde \Phi (x, \theta, \bar\theta) +
\partial_\theta \tilde \beta^{(r)} (x, \theta) = 0 \Rightarrow
b_3 (x) = 0,\; \quad f_3 (x) + \bar f_1 (x) = 0, \nonumber\\
&&\partial_{\theta} \tilde \Phi (x, \theta, \bar\theta) +
\partial_{\bar\theta} \tilde {\bar \beta}^{(r)} (x, \bar\theta)
= 0 \Rightarrow b_3 (x) = 0,\; \quad \bar f_3 (x) + f_2 (x) = 0.
\end{array}\eqno (3.7)
$$ The above equation shows that the secondary fields of
the superfields $\tilde \Phi (x,\theta, \bar\theta)$, in the
expansion (3.1), are connected with the secondary fields of (3.6).

The stage is set now to make a judicious choice so that the
conditions in (3.5) and (3.7) could be satisfied. The following
choices for the secondary fields, in terms of the auxiliary fermionic 
fields, satisfy the required conditions, namely;
$$
\begin{array}{lcl}
\bar f_3 (x) = \rho (x) = - f_2 (x),\; \qquad f_3 (x) = \lambda (x)
= - \bar f_1 (x).
\end{array}\eqno (3.8)
$$ These lead to the following expansions of the appropriate superfields $$
\begin{array}{lcl}
\tilde \beta^{(r)} (x, \theta) &=& \beta (x) + \theta\; (- \lambda
(x)) \equiv \beta (x) + \theta\; (s_{ab} \beta (x)), \nonumber\\
\tilde {\bar \beta}^{(r)} (x, \bar\theta) &=&  \bar\beta (x) +
\bar \theta\; (- \rho (x)) \equiv  \bar\beta (x) + \bar \theta\;
(s_{b} \bar \beta (x)), \nonumber\\ \tilde \Phi^{(r)} (x,
\theta, \bar\theta) &=& \phi (x) + \theta\; (\rho (x)) + \bar
\theta\; (\lambda (x)) \equiv \phi (x) + \theta\; (s_{ab} \phi
(x)) + \bar\theta\; (s_b \phi (x)),
\end{array}\eqno (3.9)
$$ where $s_{(a)b}$ are the off-shell nilpotent (anti-)BRST
symmetry transformations quoted in (2.6) and (2.7). Thus, we have
been able to derive the (anti-)BRST symmetry transformations
associated with the local fields $\beta (x), \bar\beta (x)$ and
$\phi (x)$ of the Lagrangian densities (2.4) and (2.5) in the framework of
the superfield approach to BRST formalism [3-6,14-24]. Furthermore, the
above discussion (along with equation (2.9)) provides a glimpse 
of the mappings: $s_b \leftrightarrow
\mbox{Lim}_{\theta \to 0} (\partial/\partial \bar \theta) \leftrightarrow Q_b, 
s_{ab} \leftrightarrow
\mbox{Lim}_{\bar\theta \to 0} (\partial/\partial \theta) \leftrightarrow Q_{ab}$.

It is worth emphasizing that one would have started with the
explicit presence of the auxiliary fields $B_\mu, \rho$ and
$\lambda$ in the expansion (3.1) itself as has been the case with
the earlier works on superfield approach to BRST formalism in the
context of (non-)Abelian 1-form gauge theories (see, e.g. [3-6]
for details). However, just for the sake of generality, we have
started out with an expansion of the superfields (cf. (3.1)) which
looks quite general in nature. From the above equation (3.9), it
is clear that $s_{ab} \bar\beta = 0, s_b \beta = 0  $ and $s_b
s_{ab} \phi = 0$ because these are the coefficients of
$\theta, \bar\theta$ and $\theta\bar\theta$ in the superfield
expansion.

Let us focus on the conditions $\partial_\mu \tilde \beta^{(r)} +
\partial_{\bar\theta} \tilde {\cal F}_\mu = 0$ and
 $\partial_\mu \tilde {\bar \beta}^{(r)} +
\partial_{\theta} \tilde {\bar {\cal F}}_\mu$ $= 0$. These
requirements imply the following relationships:
$$
\begin{array}{lcl}
B_\mu^{(1)}  = -\; \partial_\mu \beta, \qquad f_\mu^{(1)} = + i\;
\partial_\mu \lambda, \qquad \bar B_\mu^{(2)}  = -\;
\partial_\mu \bar \beta, \qquad \bar f_\mu^{(2)} = - i\;
\partial_\mu \rho.
\end{array}\eqno(3.10 )
$$ The substitution of the above values in the expansions of the
superfields $\tilde {\cal F}_\mu$ and $\tilde {\bar {\cal F}}_\mu$
(cf. (3.1)), along with the identifications
$\bar B_\mu^{(1)} = \bar B_\mu$ and $B_\mu^{(2)} = - B_\mu$,
leads to the following version of their reduced form
$$
\begin{array}{lcl}
\tilde {\cal F}_\mu^{(r)} (x, \theta, \bar\theta) & = & C_\mu (x)
+ \theta\; \bar B_\mu (x) + \bar\theta\; (- \partial_\mu
\beta (x)) + \theta \; \bar\theta\; (- \partial_\mu \lambda (x))
\nonumber\\ 
&\equiv& C_\mu (x) + \theta\; (s_{ab} C_\mu (x)) + \bar\theta\;
(s_b C_\mu (x)) + \theta \; \bar\theta\; (s_b s_{ab} C_\mu (x)),
\nonumber\\
\tilde {\bar {\cal F}}^{(r)}_\mu (x, \theta,
\bar\theta) & = & \bar C_\mu (x) + \theta\; (- \partial_\mu \bar
\beta (x)) + \bar\theta\; (- B_\mu (x) + \theta \; \bar\theta
\; (+ \partial_\mu \rho (x)) \nonumber\\
&\equiv& \bar C_\mu (x) + \theta\; (s_{ab} \bar C_\mu (x)) + \bar\theta\;
(s_b \bar C_\mu (x)) + \theta \; \bar\theta\; (s_b s_{ab} \bar C_\mu (x)).
\end{array}\eqno(3.11)
$$ 
It will be noted that

(i) the above transformations $s_{(a)b}$ are from
(2.6) and (2.7) that are absolutely anticommuting and off-shell
nilpotent of order two,

(ii) the choices $\bar B^{(1)}_\mu = \bar B_\mu$
and $B_\mu^{(2)} = - B_\mu$, in terms of the auxiliary fields, is
allowed within the framework of the superfield formulation, and

(iii) the above expansion is consistent with the mappings
$s_b \leftrightarrow
\mbox{Lim}_{\theta \to 0} (\partial/\partial \bar \theta) \leftrightarrow Q_b, 
s_{ab} \leftrightarrow
\mbox{Lim}_{\bar\theta \to 0} (\partial/\partial \theta) \leftrightarrow Q_{ab}$
which states that the charges
$Q_{(a)b}$ are like the  translational generators
along the Grassmannian directions of the (4, 2)-dimensional supermanifold.

The next restriction $\partial_\mu
\tilde \Phi^{(r)} +
\partial_\theta \tilde {\cal F}^{(r)}_\mu +
\partial_{\bar\theta} \tilde {\bar {\cal F}}^{(r)}_\mu = 0$
implies the following relationship$$
\begin{array}{lcl}
\bar B_\mu (x) - B_\mu (x) + \partial_\mu \phi (x)
= 0, \end{array}\eqno(3.12)
$$ where the expansions from (3.9) and
(3.11) have been inserted into the above restriction. The above
condition is the Curci-Ferrari type restriction (cf. (2.8))
that has been invoked in the proof of the anticommutativity of the (anti-)BRST
symmetry transformations (2.6) and (2.7).  
Furthermore, it
can be noted that $\partial_\theta [\partial_\mu
\tilde \Phi^{(r)} +
\partial_\theta \tilde {\cal F}^{(r)}_\mu +
\partial_{\bar\theta} \tilde {\bar {\cal F}}^{(r)}_\mu] = 0, 
\partial_{\bar\theta} [\partial_\mu
\tilde \Phi^{(r)} +
\partial_\theta \tilde {\cal F}^{(r)}_\mu +
\partial_{\bar\theta} \tilde {\bar {\cal F}}^{(r)}_\mu] = 0$. This
observation (in view of $s_b \leftrightarrow \partial_{\bar\theta}$
and $s_{ab} \leftrightarrow \partial_\theta $), ultimately, implies that the above condition 
(3.12) is an (anti-)BRST invariant relationship (i.e.
$s_{(a)b} [\bar B_\mu (x) - B_\mu (x) + \partial_\mu \phi (x)] = 0$)
and, therefore, very much physical (in some sense).
Thus, it is the superfield approach to BRST formalism which provides the basis
for the {\it existence} and (anti-)BRST {\it invariance} of the restriction (2.8). 
The geometrical origin of (2.8), in the language of gerbes, has already 
been discussed in our earlier work [35].

We are now well prepared to concentrate on the restrictions
$\partial_{\bar\theta} \tilde {\cal B}_{\mu\nu} + \partial_\mu \tilde
 {\cal F}_\nu^{(r)} - \partial_\nu \tilde  {\cal
F}^{(r)}_\mu = 0$ and $\partial_{\theta} \tilde {\cal B}_{\mu\nu} +
\partial_\mu \tilde {\bar {\cal F}}_\nu^{(r)} - \partial_\nu \tilde
{\bar {\cal F}}_\mu^{(r)} = 0$. The insertion of the super
expansions in (3.11) and (3.1) leads to the following
relationships $$
\begin{array}{lcl}
R_{\mu\nu} &=& - (\partial_\mu C_\nu - \partial_\nu C_\mu),
\qquad \bar R_{\mu\nu} = - (\partial_\mu \bar C_\nu - \partial_\nu
\bar C_\mu), \nonumber\\ S_{\mu\nu} &=& - i (\partial_\mu \bar
B_\nu - \partial_\nu \bar B_\mu \equiv - i
(\partial_\mu B_\nu - \partial_\nu  B_\mu).
\end{array}\eqno(3.13)
$$ It is clear that the last entry in the above
equation is automatically satisfied due to the relationship given
in (3.12). The next restriction is the final restriction which enables us to
compare the l.h.s. and r.h.s. of the horizontality condition as
given in  (3.4). It is clear that the following relationships
would emerge from this (cf. (3.4))  equality: $$
\begin{array}{lcl}
\partial_\mu R_{\nu\kappa} + \partial_\nu R_{\kappa\mu} +
\partial_\kappa R_{\mu\nu} &=& 0, \qquad
\partial_\mu \bar R_{\nu\kappa} + \partial_\nu \bar R_{\kappa\mu}
+ \partial_\kappa \bar R_{\mu\nu} = 0, \nonumber\\
\partial_\mu S_{\nu\kappa} + \partial_\nu S_{\kappa\mu} + \partial_\kappa
S_{\mu\nu} &=& 0.
\end{array}\eqno(3.14)
$$ These conditions are readily satisfied by the values
obtained for the expressions of the secondary fields $R_{\mu\nu},
\bar R_{\mu\nu}$ and $S_{\mu\nu}$ in terms of the basic and auxiliary fields
(cf. (3.13)). As a consequence, we note that the super curvature tensor
$\tilde H_{\mu\nu\kappa}^{(h)} = \partial_\mu {\cal B}^{(h)}_{\nu\kappa} +
\partial_\nu {\cal B}^{(h)}_{\kappa\mu} + \partial_\kappa {\cal B}^{(h)}_{\mu\nu}
\equiv H_{\mu\nu\kappa}$ 
remains independent of the Grassmannian variables $\theta$ and $\bar\theta$
(because $\tilde {\cal B}^{(h)}_{\mu\nu} (x, \theta, \bar\theta) =
B_{\mu\nu} (x) - \theta (\partial_\mu \bar C_\nu - \partial_\nu \bar C_\mu) -
 \bar\theta (\partial_\mu  C_\nu - \partial_\nu  C_\mu)
+ \theta \bar\theta\; (\partial_\mu  B_\nu - \partial_\nu  B_\mu)$) where 
the superscript $(h)$ on $\tilde H^{(h)}_{\mu\nu\kappa}$ denotes that the
super curvature tensor has been obtained after the application of the 
horizontality condition due to which ${\cal B}_{\mu\nu} (x,\theta,\bar\theta)
\to  {\cal B}_{\mu\nu}^{(h)} (x,\theta,\bar\theta)$.

The substitution of all the above values of the secondary fields,
in terms of the auxiliary and basic fields, leads to the following
expansion for (3.1), namely;
 $$
\begin{array}{lcl}
\tilde {\cal B}^{(h)}_{\mu\nu} (x, \theta, \bar\theta) &=&
B_{\mu\nu} (x) + \theta\; (s_{ab} B_{\mu\nu} (x)) + \bar\theta\;
(s_b B_{\mu\nu} (x)) + \theta\; \bar\theta\; (s_b s_{ab}
B_{\mu\nu} (x)), \nonumber\\ \tilde \beta^{(h)} (x, \theta,
\bar\theta ) &=& \beta (x) + \theta \;(s_{ab} \beta (x)) +
\bar\theta\; (s_b \beta (x)) + \theta\; \bar\theta\; (s_b s_{ab}
\beta (x)), \nonumber\\ \tilde {\bar \beta}^{(h)} (x, \theta,
\bar\theta) &=&  \bar\beta (x) + \theta \; (s_{ab} \bar \beta
(x)) + \bar \theta\; (s_b \bar \beta (x)) + \theta\; \bar\theta\;
(s_b s_{ab} \bar \beta (x)), \nonumber\\ \tilde \Phi^{(h)} (x,
\theta, \bar\theta) &=& \phi (x) + \theta \;(s_{ab} \phi (x))
+ \bar\theta\; (s_b \phi (x)) + \theta\; \bar\theta \; (s_b
s_{ab} \phi (x)),  \nonumber\\ \tilde {\cal F}^{(h)}_\mu (x,
\theta, \bar\theta) &=& C_\mu (x) + \theta \;(s_{ab} C_\mu (x)) +
\bar\theta\; (s_b C_\mu (x)) + \theta\; \bar\theta\; (s_b s_{ab}
C_\mu (x)), \nonumber\\ \tilde {\bar {\cal F}}^{(h)}_\mu (x,
\theta, \bar\theta) &=& \bar C_\mu (x) + \theta\; (s_{ab} \bar
C_\mu (x)) + \bar\theta\; (s_b \bar C_\mu (x)) + \theta\;
\bar\theta\;  (s_b s_{ab} \bar C_\mu(x)).
\end{array}\eqno (3.15)
$$ Here (i) the off-shell nilpotent transformations $s_{(a)b}$ of
equations (2.6) and (2.7) have been taken into account for the
above expansions, and (ii) the superscript $(h)$ on the above
superfields denotes the expansion of the superfields after the
application of the horizontality condition. Furthermore, it will
be noted that, in the above expansion, we have taken into account
$s_b \beta = 0, s_{ab} \bar \beta = 0, s_b s_{ab} \phi \equiv
s_{ab} s_b \phi = 0$. Finally, the geometrical interpretations for the
off-shell nilpotent (anti-)BRST symmetry transformations and their
corresponding charges emerge from the following relationships (cf. (2.9)) 
$$
\begin{array}{lcl}
&&\mbox{Lim}_{\bar\theta \to 0}\; {\displaystyle
\frac{\partial}{\partial\theta}} \; \tilde \Omega^{(h)} (x, \theta,
\bar\theta) = s_{ab} \Omega (x) \equiv - i [\Omega (x),
Q_{ab}]_{(\pm)}, \nonumber\\ &&\mbox{Lim}_{\theta \to 0}\;
{\displaystyle \frac{\partial}{\partial\bar\theta}} \; \tilde
\Omega^{(h)} (x, \theta, \bar\theta) = s_{b} \Omega (x) \equiv - i
[\Omega (x), Q_{b}]_{(\pm)},
\end{array}\eqno(3.16)
$$ where $\Omega (x)$ is the generic local field of the Lagrangian
densities (2.4)/(2.5) and $\tilde \Omega^{(h)} (x, \theta, \bar\theta)$ is the
corresponding superfield defined on the (4, 2)-dimensional
supermanifold (cf. (3.15)).

The above expression implies that the off-shell nilpotent
(anti-)BRST symmetry transformations $s_{(a)b}$ and their
corresponding generators $Q_{(a)b}$ geometrically correspond to
the translational generators along the Grassmannian directions of
the (4, 2)-dimensional supermanifold. To be more specific, the
BRST symmetry transformation corresponds to the translation of the
particular superfield along the $\bar\theta$-direction of the
supermanifold when there is no translation of the same superfield
along the $\theta$-direction of the supermanifold (i.e. $\theta
\to 0$). This geometrical operation on the specific superfield
generates the BRST symmetry transformation for the corresponding
4D ordinary field present in the Lagrangian density (2.4). A
similar kind of argument can be provided for the existence of the
anti-BRST symmetry transformation for a specific field in the
language of the translational generator 
(i.e. $ \mbox{Lim}_{\bar\theta \to 0}\; (\partial/\partial\theta)$)
on the above (4, 2)-dimensional supermanifold.

It will be noted that the (anti-)BRST symmetry transformations for
the auxiliary fields ($B_\mu, \bar B_\mu$) and the Lagrange multiplier field
($L_\mu$) are derived from the requirement of the absolute anticommutativity 
$[s_b s_{ab} + s_{ab} s_b]\; \Omega = 0$ of the (anti-)BRST symmetry transformations
$s_{(a)b}$ for  a generic field $\Omega$. For instance, it can be seen, 
from (2.6) and (2.7), that
$(s_b s_{ab} + s_{ab} s_b)\; C_\mu = 0$ and 
$(s_b s_{ab} + s_{ab} s_b) \;\bar C_\mu = 0$ yield the nilpotent
transformations $s_b \bar B_\mu = - \partial_\mu \lambda$ and $
s_{ab} B_\mu = \partial_\mu \rho$, respectively. Similarly, the nilpotent
transformations for the multiplier field $L_\mu$ are found from the equations
of motion $L_\mu = \bar B_\mu$ and $L_\mu = - B_\mu$ that emerge from the
Lagrangian densities (2.4) and (2.5), respectively. It is elementary now to
note that $s_b L_\mu = - \partial_\mu \lambda$ and 
$s_{ab} L_\mu = - \partial_\mu \rho$. Thus, ultimately, we obtain all
the nilpotent (anti-)BRST symmetry transformations for all the fields
of the theory.

\section{(Anti-)BRST invariance: superfield approach}

\noindent 
The nilpotent and absolutely anticommuting (anti-)BRST symmetry 
transformations (2.6) and (2.7) leave the Lagrangian densities
(2.4) and (2.5) quasi-invariant because the latter transform
to the total spacetime derivatives. This observation can be 
captured within the framework of our present superfield approach 
to BRST formalism. To this end in mind, let us, first of all, 
note that the following relationship [35] is true, namely;
$$
\begin{array}{lcl}
&& s_b\; s_{ab}\; \Bigl [ \;2 \beta \bar\beta + \bar C_\mu C^\mu - 
{\displaystyle \frac{1}{4}}
B^{\mu\nu} B_{\mu\nu} \;\Bigr ] = B^\mu (\partial^\nu B_{\nu\mu})
+ B \cdot \bar B + \partial_\mu \bar \beta  \partial^\mu \beta\\
&& + (\partial_\mu \bar C_\nu - \partial_\nu \bar C_\mu) (\partial^\mu C^\nu)
+ (\partial \cdot C - \lambda) \; \rho + (\partial \cdot \bar C + \rho)\; \lambda.
\end{array}\eqno (4.1)
$$
In the above, using (2.8)/(3.12), we can re-express
$$
\begin{array}{lcl}
B \cdot \bar B = {\displaystyle \frac{1}{2} \Bigl ( B \cdot B + \bar B \cdot \bar B
\Bigr ) - \frac{1}{2} \partial^\mu \phi \partial_\mu \phi}, 
\end{array}\eqno (4.2)
$$
to obtain the gauge-fixing and Faddeev-Popov ghost terms of the Lagrangian densities
(2.4) and (2.5). This shows that, modulo some total spacetime derivatives terms, the gauge-fixing
and Faddeev-Popov ghost terms are actually the (anti-)BRST exact terms.

The horizontality condition leads to the explicit expression of the 2-form gauge superfield,
in terms of the basic and auxiliary fields,  as
 $$
\begin{array}{lcl}
\tilde {\cal B}^{(h)}_{\mu\nu} (x, \theta, \bar\theta) =
B_{\mu\nu} (x) - \theta \;(\partial_\mu \bar C_\nu - \partial_\nu \bar C_\mu) -
 \bar\theta \;(\partial_\mu  C_\nu - \partial_\nu  C_\mu)
+ \theta \bar\theta\; (\partial_\mu  B_\nu - \partial_\nu  B_\mu),
\end{array}\eqno (4.3)
$$
where the last term can also be expressed as $ \theta \bar\theta\; (\partial_\mu  \bar B_\nu - \partial_\nu  \bar B_\mu)$ (cf. (3.13)). Using the explicit expansions in (3.9), (3.11) and (4.3),
it can be seen that the Lagrangian densities (2.4) and (2.5) can be expressed, in terms
of the superfields, as
$$
\begin{array}{lcl}
\tilde {\cal L}_B^{(b, ab)} &=& {\displaystyle \frac{1}{12} \tilde H^{\mu\nu\kappa (h)} \tilde H^{(h)}_{\mu\nu\kappa} + \frac{\partial}{\partial \bar\theta}\; \frac{\partial}{\partial \theta}
\Bigl [ 2 \;\tilde \beta^{(r)} \;\tilde {\bar \beta}^{(r)} 
+ \tilde {\bar {\cal F}}^{(r)} \cdot \tilde {\cal F}^{(r)} 
- \frac{1}{4}\; \tilde {\cal B}^{(h)}_{\mu\nu}\; \tilde {\cal B}^{(h)\mu\nu} 
\Bigr ]} \nonumber\\ &-& L^\mu (x) \;\Bigl [ \partial_\mu \tilde \Phi^{(r)} + \partial_\theta \tilde
{\cal F}^{(r)}_\mu + \partial_{\bar\theta} \tilde {\bar {\cal F}}^{(r)}_\mu \Bigr ],
 \end{array}\eqno (4.4)
$$
where $\tilde H_{\mu\nu\kappa}^{(h)} = \partial_\mu {\cal B}^{(h)}_{\nu\kappa} +
\partial_\nu {\cal B}^{(h)}_{\kappa\mu} + \partial_\kappa {\cal B}^{(h)}_{\mu\nu}$ 
is the curvature
super tensor derived after the application of the horizontality condition. It is straightforward to note that this super tensor is independent of the Grassmannian variables. As a consequence, the
kinetic energy term for the 2-form gauge field is an (anti-)BRST invariant quantity (i.e.
$s_{(a)b} H_{\mu\nu\kappa} = 0$).

Taking the help of discussions after equation (3.12) and exploiting the nilpotency property
of the translational generators (i.e. $\partial_\theta^2 = 0, \partial_{\bar\theta}^2 = 0$), 
it is evident that the following relationship is sacrosanct, namely;
$$
\begin{array}{lcl}
&& \mbox{Lim}_{\theta \to 0} \;{\displaystyle \frac{\partial}{\partial \bar \theta}} \tilde
{\cal L}_B^{(b, ab)} = 0 \qquad \Leftrightarrow \qquad s_b {\cal L}_B^{(b)} = 0, \nonumber\\ 
&& \mbox{Lim}_{\bar\theta \to 0} \;{\displaystyle \frac{\partial}{\partial \theta}} \tilde
{\cal L}_B^{(b, ab)} = 0 \qquad \Leftrightarrow \qquad s_{ab} {\cal L}_B^{(ab)} = 0.
\end{array}\eqno(4.5)
$$
Thus, we conclude that the Grassmannian independence of the super Lagrangian density,
expressed in terms of the (4, 2)-dimensional superfields (derived after the application
of the horizontality condition), provides a clear-cut proof for the (anti-)BRST invariance of the
4D Lagrangian densities (2.4) and (2.5). In other words, if the operation of the
partial derivatives w.r.t. the Grassmannian variables, on the appropriate (4, 2)-dimensional 
super Lagrangian density, turns out to be zero, the corresponding 4D Lagrangian density
of a given gauge theory
would respect the nilpotent (anti-)BRST symmetry invariance. This conclusion is in complete agreement with our recent works on 1-form gauge theories [36-39].

We wrap up this section with the assertion that the superfield approach to BRST formalism
does simplify the understanding of the (anti-)BRST invariance in a given theory.

\section{Conclusions}

\noindent In our present endeavour, we have concentrated on the
application of the geometrical superfield approach to BRST
formalism to derive the off-shell 
nilpotent and absolutely anticommuting
(anti-)BRST symmetry transformations for {\it all} the
fields of the Lagrangian densities (cf. (2.4),(2.5))
 of a 4D free Abelian 2-form gauge
theory. To the best of our knowledge, this is for the first time
that the idea of the geometrical superfield approach to BRST formalism
(especially proposed in [3-6, 14-24] for the 4D
(non-)Abelian 1-form gauge theories) has been generalized to the case
of the free 4D Abelian 2-form gauge theory. The above geometrical
superfield approach, we firmly believe, can be extended so as to
derive the (anti-)BRST symmetry transformations in the case
of the higher p-form ($p \geq 3$) gauge theories which have become important 
in the context of the (super)string theories.

Our present study illustrates that there is {\it no}
existence of an absolutely anticommuting set of
 {\it on-shell} nilpotent (anti-)BRST symmetry 
transformations for the 4D free Abelian 2-form gauge theory. 
This feature of our present Abelian gauge theory is exactly same as that
of the 4D non-Abelian 1-form gauge theory where the on-shell nilpotent and anticommuting
(anti-)BRST symmetry transformations do not exist {\it together} [9-12]. The off-shell nilpotent
and absolutely anticommuting (anti-)BRST transformations exist
for both the above theories due to the CF type restrictions
which emerge from a pair of coupled Lagrangian densities. Thus, our present free
Abelian 2-form theory does imbibe some of the key features of the non-Abelian
1-form gauge theory.

It will be noted that, in our very recent work [40], we have been able to show 
the existence of the on-shell and off-shell nilpotent BRST symmetry transformations
for a specific Lagrangian density for the 4D free Abelian 2-form gauge theory. In
a similar fashion, for another specific Lagrangian density of the above 2-form gauge theory,
the on-shell and off-shell nilpotent anti-BRST symmetry transformations have also been shown
to exist. However, the on-shell nilpotent (anti-)BRST symmetry transformations do not
exist {\it together} for a single Lagrangian density of the above 2-form gauge theory.

We note that there is a great
deal of difference between the 4D Abelian 1-form gauge theory (that is endowed 
with the off-shell as well as on-shell nilpotent and anticommuting (anti-)BRST symmetries) and
the 4D free Abelian 2-form gauge theory. The latter theory has deep connection 
with the geometrical objects called gerbes [35] (due to the restriction (2.8)/(3.12)
which is not the case with the 4D Abelian 1-form gauge theory where there
is no need of any CF type restriction). In fact, for the Abelian 1-form theory,
the (anti-)BRST symmetry transformations are found to be automatically anticommuting.

One can encapsulate the
geometrical interpretations (see, e.g., [14-24] for details)
of specific quantities, connected with the BRST formalism, in the language of the
following mathematical mappings  $$
\begin{array}{lcl}
&&s_b \Leftrightarrow Q_b \;\Leftrightarrow \;\mbox{Lim}_{\theta \to
0} \;{\displaystyle \frac{\partial}{\partial\bar\theta}}, \qquad
s_{ab} \Leftrightarrow Q_{ab} \;\Leftrightarrow\;
\mbox{Lim}_{\bar\theta \to 0} \;{\displaystyle
\frac{\partial}{\partial\theta}}, \nonumber\\ && s_b^2 = 0
\Leftrightarrow Q_b^2 = 0 \;\;\;\Leftrightarrow \;\;\;\mbox{Lim}_{\theta \to
0}\; \Bigl ( {\displaystyle \frac{\partial}{\partial\bar\theta}}
\Bigr )^2 = 0, \nonumber\\ && s_{ab}^2 = 0 \;\;\;\Leftrightarrow \;\;\;Q_{ab}^2
= 0 \;\;\Leftrightarrow \;\;\mbox{Lim}_{\bar\theta \to 0}\;\Bigl (
{\displaystyle \frac{\partial}{\partial\theta}} \Bigr )^2 = 0, \nonumber\\
&& s_b s_{ab} + s_{ab} s_b = 0 \;\;\;\Leftrightarrow \;\;\;Q_b Q_{ab} + Q_{ab} Q_b = 0 
\;\;\;\Leftrightarrow \;\;\;\nonumber\\
&& \Bigl (\mbox{Lim}_{\bar\theta \to 0} \;{\displaystyle
\frac{\partial}{\partial\theta}} \Bigr ) \; \Bigl (
\mbox{Lim}_{\theta \to
0} \;{\displaystyle \frac{\partial}{\partial\bar\theta}} \Bigr )
+ \Bigl (\mbox{Lim}_{\theta \to
0} \;{\displaystyle \frac{\partial}{\partial\bar\theta}} \Bigr )\;
\Bigl (\mbox{Lim}_{\bar\theta \to 0} \;{\displaystyle
\frac{\partial}{\partial\theta}} \Bigr ) = 0.
\end{array}\eqno(5.1)
$$ The above (geometrically intuitive) mappings are possible only in
the super field approach to BRST formalism proposed in [3-6,14-24].
This is not the case, however, with the mathematical superfield
approach to BRST formalism proposed in [41,42].

It is clear from the above equation (cf. (5.1)) that the BRST and anti-BRST charges 
have their own identity and they play completely independent
roles. In fact, they correspond to the translational generators along the independent
Grassmannian directions $\theta$ and $\bar\theta$ of the (4, 2)-dimensional supermanifold
on which the present Abelian 2-form gauge theory is considered. A clear-cut proof of the above
assertion has been corroborated in our earlier work [40] where it has been demonstrated
that the BRST and anti-BRST charges lead to completely distinct and independent constraint
conditions. These conditions emerge from the physicality criteria $Q_{(a)b} |phys> = 0$ [40].
The latter is exploited in the context of
discussions connected with the constraint structure of the 4D Abelian 2-form gauge theory.

We touch upon another very decisive feature of the present superfield 
approach to BRST formalism. This formulation always ensures the nilpotency
and an absolute anticommutativity of the (anti-)BRST symmetry transformations
as is evident from the last two entries in the above equation (5.1). In fact,
it is due to the above key points that we obtain the CF type 
restriction (cf. (3.12)) from the superfield approach to BRST formalism
which ensures the anticommutativity of the (anti-)BRST symmetry transformations
$s_{(a)b}$. The validity of $s_b s_{ab} + s_{ab} s_b = 0$ turns out explicitly
from the super expansion (3.15) because it can be noted that, for all the
superfields $\tilde \Omega^{(h)}$
(derived after the application of the horizontality condition),
the operator equation $[(\partial/\partial \theta) (\partial/\partial \bar \theta)
+ (\partial/\partial \bar \theta) \partial/\partial \theta)]\;
 \tilde \Omega^{(h)} (x,\theta,\bar\theta) = 0$ is always true. Thus, one of the 
key results of our present investigation is 
the derivation of the absolutely anticommuting 
(anti-)BRST symmetry transformations.

It will be noted that the horizontality condition $\tilde d \tilde
B^{(2)} = d B^{(2)}$ is a gauge (i.e. nilpotent (anti-)BRST) invariant
restriction on an appropriately chosen super 2-form
gauge connection $\tilde B^{(2)}$ that is
defined on a suitably selected supermanifold. This is due to the
fact that the curvature tensor $H_{\mu\nu\kappa}$, that
constitutes the 3-form $H^{(3)} = d B^{(2)}$, remains invariant
under the (anti-)BRST symmetry transformations (i.e. $s_{(a)b}
H_{\mu\nu\kappa} = 0$). As commented earlier after equation (2.6),
the key reasons behind the emergence of the nilpotent (anti-)BRST
symmetry transformations {\it together}, within the framework of
the superfield formulation, is encoded (i) physically in the
observation that $s_{(a)b} H_{\mu\nu\kappa} = 0$, and (ii)
mathematically in the nilpotency $\tilde d^2 = 0$ of the super
exterior derivative ($\tilde d = dx^\mu \partial_\mu + d\theta
\partial_\theta + d \bar\theta \partial_{\bar\theta}$).

It has been shown in our earlier works [32-34] that, in addition
to the above nilpotent (anti-)BRST symmetry transformations, there
exist nilpotent (anti-)co-BRST symmetry transformations for the
Abelian 2-form gauge theory. However, these transformations have been found
to be anticommuting only up to a U(1) vector gauge transformation.   
In our very recent works [43,44], we have derived the absolutely anticommuting
set of (anti-)BRST as well as (anti-)co-BRST symmetry transformations and shown
that the Abelian 2-form gauge theory is a field theoretic model for the Hodge theory.
It would be very interesting to extend our present work and exploit the
superfield approach to derive the above absolutely anticommuting
(anti-)co-BRST symmetry transformations for the theory.

To generalize our present idea to the non-Abelian 2-form gauge theory
is another promising direction. Yet
another direction, that could be pursued for the application of
the superfield approach to BRST formalism, is in the context of
interesting field theoretical models proposed in [45,46] which
also involve the 2-form gauge field. These are some of the issues
that are presently being investigated and our results
would be reported elsewhere [47].\\

\begin{center}
{\bf Appendix A}
\end{center}

Our present superfiled formalism can be applied
to any arbitrary Abelian
$p$-form gauge theory in any arbitrary $D$-dimension of spacetime.
To corroborate this assertion, we apply our method first
to the 4D Abelian 3-form 
($B^{(3)} = \frac{1}{3!} (dx^\mu \wedge dx^\nu \wedge dx^\eta) B_{\mu\nu\eta}$)
gauge theory described by a totally antisymmetric potential $B_{\mu\nu\eta}$.
The generalization of the above ordinary 4D 3-form 
onto the  (4, 2)-dimensional supermanifold 
(with $Z^M = (x^\mu, \theta, \bar\theta)$) is 
$$
\begin{array}{lcl}
&& B^{(3)} \to \tilde B^{(3)} = {\displaystyle \frac{1}{3!}\; (d Z^M \wedge d
Z^N \wedge d Z^K) \tilde B_{MNK} \equiv \frac{1}{3!}\; (dx^\mu \wedge dx^\nu \wedge dx^\eta)\;
\tilde B_{\mu\nu\eta}} \nonumber\\ && + 
{\displaystyle \frac{1}{2} (dx^\mu \wedge dx^\nu \wedge d\theta) \tilde
B_{\mu\nu\theta} + \frac{1}{2} (dx^\mu \wedge dx^\nu \wedge d \bar\theta)
\tilde B_{\mu\nu\bar\theta} + \frac{1}{3!} (d\theta \wedge d\theta
\wedge d\theta) \tilde B_{\theta\theta\theta}} \nonumber\\ &&+  
{\displaystyle \frac{1}{3!}\; (d
\bar\theta \wedge d\bar\theta \wedge d \bar\theta) \;\tilde B_{\bar\theta\bar\theta\bar\theta} +
(dx^\mu \wedge d \theta \wedge d\bar\theta) \tilde B_{\mu\theta\bar\theta} + \frac{1}{2}
(dx^\mu \wedge d\theta \wedge d \theta) \tilde B_{\mu\theta\theta}} \nonumber\\
&& + {\displaystyle \frac{1}{2} (dx^\mu \wedge d\bar\theta \wedge d \bar\theta) \tilde B_{\mu\bar\theta\bar\theta}
+ \frac{1}{2} (d\theta \wedge d\bar\theta \wedge d \bar\theta) \;\tilde B_{\theta\bar\theta\bar\theta}
+ \frac{1}{2} (d\theta \wedge d\theta \wedge d \bar\theta) \;\tilde B_{\theta\theta\bar\theta} }.
\end{array}\eqno(A.1)
$$
We make the identifications: $\tilde B_{\mu\nu\eta} = \tilde {\cal B}_{\mu\nu\eta} 
(x, \theta,\bar\theta)$, $\tilde B_{\mu\nu\theta} =
\tilde {\bar {\cal F}}_{\mu\nu} (x, \theta, \bar\theta)$, $\tilde
B_{\mu\nu\bar\theta} = \tilde {\cal F}_{\mu\nu} (x,\theta,\bar\theta)$,
$\tilde B_{\mu\theta\bar\theta} = \tilde \Phi_\mu
(x,\theta,\bar\theta)$, $\frac{1}{3!} \tilde B_{\theta\theta \theta} = \tilde
{\bar {\cal F}}_2 (x,\theta,\bar\theta)$, $\frac{1}{3!}
B_{\bar\theta\bar\theta\bar\theta} = \tilde {\cal F}_2  (x,\theta,\bar\theta)$,
$\frac{1}{2} \tilde B_{\theta\bar\theta\bar\theta} = \tilde {\cal F}_1
(x,\theta,\bar\theta)$, $\frac{1}{2} \tilde B_{\theta\theta\bar\theta} = \tilde 
{\bar {\cal F}}_1 (x,\theta,\bar\theta)$, $\frac{1}{2} \tilde B_{\mu\bar\theta\bar\theta}
= \tilde \beta_\mu (x, \theta, \bar\theta)$ and $\frac{1}{2} \tilde B_{\mu\theta\theta}
= \tilde {\bar \beta}_\mu (x, \theta, \bar\theta)$ as the generalization of the 4D local
fields $(B_{\mu\nu\eta}, \bar C_{\mu\nu}, C_{\mu\nu}, \phi_\mu, \bar C_2, C_2, \bar C_1, C_1,
\beta_\mu, \bar\beta_\mu )$ 
of the Abelian 3-form gauge theory onto the corresponding superfields 
defined on the (4, 2)-dimensional supermanifold.

Our present Abelian 3-form
gauge theory has to be considered on this (4, 2)-dimensional
supermanifold in the framework of
the superfield approach to BRST formalism so that we can derive the nilpotent
and absolutely anticommuting (anti-)BRST symmetry transformations for all
the fields of the theory by exploiting the theoretical arsenal of 
horizontality condition. Towards this goal in mind, we shall quote the
main results emerging from the application of the horizontality condition
and shall avoid the algebraic details.

The super expansion of the above superfields along the Grassmannian directions
of the supermanifold can be expressed as follows
$$
\begin{array}{lcl}
\tilde {\cal B}_{\mu\nu\eta} (x, \theta, \bar\theta) &=& B_{\mu\nu\eta}
(x) + \theta\; \bar R_{\mu\nu\eta} (x) + \bar\theta\; R_{\mu\nu\eta} (x) +
i \;\theta \; \bar\theta\; S_{\mu\nu\eta} (x), \nonumber\\ \tilde \beta_\mu
(x, \theta, \bar\theta ) &=& \beta_\mu (x) + \theta \;\bar f^{(1)}_\mu (x) +
\bar\theta\; f^{(1)}_\mu (x) + i\; \theta\; \bar\theta\; b_\mu (x),
\nonumber\\ \tilde {\bar \beta}_\mu (x, \theta, \bar\theta) &=& 
\bar\beta_\mu (x) + \theta \;\bar f^{(2)}_\mu (x) + \bar \theta\; f^{(2)}_\mu (x) 
+ i\;\theta\;\bar\theta\; \bar b_\mu (x), \nonumber\\ \tilde \Phi_\mu (x,
\theta, \bar\theta) &=& \phi_\mu (x) + \theta \;\bar f^{(3)}_\mu (x) +
\bar\theta\; f^{(3)}_\mu (x) + i \;\theta \;\bar\theta\; b^{(3)}_\mu (x),
\nonumber\\ \tilde {\cal F}_{\mu\nu} (x, \theta, \bar\theta) &=& C_{\mu\nu}
(x) + \theta \;\bar B^{(1)}_{\mu\nu} (x) + \bar\theta\; B^{(1)}_{\mu\nu} (x)
+ i \;\theta \; \bar\theta\; s_{\mu\nu} (x), \nonumber\\ \tilde
{\bar {\cal F}}_{\mu\nu} (x, \theta, \bar\theta) &=& \bar C_{\mu\nu} (x) +
\theta\; \bar B^{(2)}_{\mu\nu} (x) + \bar\theta\; B^{(2)}_{\mu\nu} (x) + i
\; \theta\; \bar\theta \;\bar s_{\mu\nu} (x), \nonumber\\
\tilde {\cal F}_1 (x, \theta, \bar\theta) &=& C_1 (x) + \theta\;\bar b_1^{(1)} (x)
+ \bar \theta\; b_1^{(1)} (x) + i\;\theta\;\bar\theta\; s_1 (x), \nonumber\\
\tilde {\bar {\cal F}}_1 (x, \theta, \bar\theta) &=& \bar C_1 (x) + \theta\;\bar b_1^{(2)} (x)
+ \bar \theta\; b_1^{(2)} (x) + i\;\theta\;\bar\theta\; \bar s_1 (x), \nonumber\\
\tilde {\cal F}_2 (x, \theta, \bar\theta) &=& C_2 (x) + \theta\;\bar b_2^{(1)} (x)
+ \bar \theta\; b_2^{(1)} (x) + i\;\theta\;\bar\theta\; s_2 (x), \nonumber\\
\tilde {\bar {\cal F}}_2 (x, \theta, \bar\theta) &=& \bar C_2 (x) + \theta\;\bar b_2^{(2)} (x)
+ \bar \theta\; b_2^{(2)} (x) + i\;\theta\;\bar\theta\; \bar s_2 (x), 
\end{array}\eqno (A.2)
$$ where $B_{\mu\nu\eta}$ is the gauge field, $\phi_\mu$ is a vector bosonic field,
$(\bar C_{\mu\nu}) C_{\mu\nu}$ are the fermionic antisymmetric (anti-)ghost fields,
$(\bar \beta_\mu)\beta_\mu$ are the bosonic vector (anti-)ghost fields, $(\bar C_2)C_2$
and $(\bar C_1)C_1$ are a pair of fermionic (anti-)ghost Lorentz scalar fields. These
fields are the basic (primary) fields of the 4D Abelian 3-form gauge theory within the framework
of the BRST formalism. Rest of the fields, in the above expansion, are the secondary fields 
that have to be expressed in terms of the primary (basic) fields.

By exploiting the horizontality condition $\tilde d \tilde B^{(3)} = d B^{(3)}$\footnote{Note
that $d B^{(3)} = \frac{1}{4!} (dx^\mu \wedge dx^\nu \wedge dx^\eta \wedge dx^\xi)
H_{\mu\nu\eta\xi}$ where $H_{\mu\nu\eta\xi} = \partial_\mu B_{\nu\eta\xi}
+ \partial_\nu B_{\eta\xi\mu} + \partial_\eta B_{\xi\mu\nu} + \partial_\xi B_{\mu\nu\eta}$ is
a totally antisymmetric curvature tensor that would be 
useful for the kinetic term of the gauge field.}, where $\tilde d$
and $\tilde B^{(3)}$ are defined in (3.2) and (A.1), one would be able to achieve the above goal
as well as one would be able to derive the nilpotent and absolutely anticommuting (anti-)BRST
symmetry transformations of the theory. The algebraic computations are a bit tedious but straightforward
like that of our Sec. 3. The results, emerging from the above horizontality condition
by setting all the Grassmannian components of the supercurvature tensor equal to zero, are
$$ 
\begin{array}{lcl}
&& b_2^{(1)} = 0, \quad s_2 = 0, \quad \bar b_2^{(2)} = 0, \quad \bar s_2 = 0, \quad
\bar s_1 = 0, \quad s_1 = 0, \quad b_2^{(2)} + \bar b_1^{(2)} = 0,  \nonumber\\
&& \bar b_2^{(1)} + b_1^{(1)} = 0, \quad b_1^{(2)} + \bar b_1^{(1)} = 0, \quad \bar f_\mu^{(2)}
= \partial_\mu \bar C_2, \quad f_\mu^{(1)} = \partial_\mu C_2, \quad 
b_\mu = i \partial_\mu \bar b_2^{(1)}, \nonumber\\
&& \bar b_\mu = - i \partial_\mu b_2^{(2)}, \quad b_\mu^{(3)} = - i \partial_\mu b_1^{(2)},
\quad B^{(1)}_{\mu\nu} = \partial_\mu \beta_\nu - \partial_\nu \beta_\mu, \quad
\bar f_\mu^{(2)} = \partial_\mu \bar C_2, \nonumber\\
&& \bar B_{\mu\nu}^{(2)} = \partial_\mu \bar \beta_\nu - \partial_\nu \bar \beta_\mu, \quad
s_{\mu\nu} = i (\partial_\mu \bar f_\nu^{(1)} - \partial_\nu \bar f_\mu^{(1)}) \equiv
- i (\partial_\mu  f_\nu^{(3)} - \partial_\nu  f_\mu^{(3)}), \nonumber\\
&& \bar s_{\mu\nu} = + i (\partial_\mu \bar f_\nu^{(3)} - \partial_\nu \bar f_\mu^{(3)}) \equiv
- i (\partial_\mu  f_\nu^{(2)} - \partial_\nu  f_\mu^{(2)}), \nonumber\\ && R_{\mu\nu\eta}
= \partial_\mu C_{\nu\eta} + \partial_\nu C_{\eta\mu} + \partial_\eta C_{\mu\nu}, \quad
 \bar R_{\mu\nu\eta}
= \partial_\mu \bar C_{\nu\eta} + \partial_\nu \bar C_{\eta\mu} + \partial_\eta \bar C_{\mu\nu},
\nonumber\\&&  S_{\mu\nu\eta} = - i (\partial_\mu B_{\nu\eta}^{(2)} + \partial_\nu B_{\eta\mu}^{(2)}
+ \partial_\eta B_{\mu\nu}^{(2)}) \equiv
 + i (\partial_\mu \bar B_{\nu\eta}^{(1)} + \partial_\nu \bar B_{\eta\mu}^{(1)}
+ \partial_\eta \bar B_{\mu\nu}^{(1)}). 
\end{array}\eqno(A.3)
$$
In addition to the above results, we obtain the following Curci-Ferrari type restrictions
$$ \begin{array}{lcl}
&& f_\mu^{(2)} + \bar f_\mu^{(3)} = \partial_\mu \bar C_1, \quad
 \bar f_\mu^{(1)} +  f_\mu^{(3)} = \partial_\mu  C_1, \quad
\bar B_{\mu\nu}^{(1)} + B_{\mu\nu}^{(2)} = \partial_\mu \phi_\nu - \partial_\nu \phi_\mu,
\end{array}\eqno(A.4)
$$
which ensure the consistency of the {\it three} equivalences shown in (A.3). It will be noted that
the above restrictions emerge from setting the specific coefficients of the 4-form differentials
[e.g. $(dx^\mu \wedge d\theta \wedge d\theta \wedge d \bar\theta),
(dx^\mu \wedge d\theta \wedge d \bar \theta \wedge d \bar\theta), 
(dx^\mu \wedge d x^\nu \wedge d\theta \wedge d \bar\theta)$] of the l.h.s. of  the horizontality condition
$\tilde d \tilde B^{(3)} = d B^{(3)}$.  Finally, it is worth pointing out that the coefficients of
the differentials $(dx^\mu \wedge dx^\nu \wedge dx^\eta \wedge dx^\xi)$ from the l.h.s. and r.h.s.
of the condition $\tilde d \tilde B^{(3)} = d B^{(3)}$ match due to the precise form
of $R_{\mu\nu\eta}, \bar R_{\mu\nu\eta} $ and $S_{\mu\nu\eta}$,  quoted in (A.3).

The substitution of the results of (A.3) into (A.2) leads to the following super expansion
of the superfields (after the application of the horizontality condition) in the language of the nilpotent
and absolutely (anti-)BRST symmetry transformations
$$
\begin{array}{lcl}
\tilde {\cal B}^{(h)}_{\mu\nu\eta} (x, \theta, \bar\theta) &=& B_{\mu\nu\eta}
(x) + \theta\; (s_{ab} B_{\mu\nu\eta} (x)) + \bar\theta\; (s_b B_{\mu\nu\eta} (x)) +
 \theta \; \bar\theta\; (s_b s_{ab} B_{\mu\nu\eta} (x)), \nonumber\\ 
\tilde \beta^{(h)}_\mu
(x, \theta, \bar\theta ) &=& \beta_\mu (x) + \theta \;(s_{ab} \beta_\mu (x)) +
\bar\theta\; (s_b \beta_\mu (x)) +  \theta\; \bar\theta\; (s_b s_{ab} \beta_\mu (x)),
\nonumber\\ \tilde {\bar \beta}^{(h)}_\mu (x, \theta, \bar\theta) &=& 
\bar\beta_\mu (x) + \theta \;(s_{ab} \bar\beta_\mu (x)) 
+ \bar \theta\; (s_b \bar\beta_\mu (x)) 
+ \theta\;\bar\theta\; (s_b s_{ab} \bar\beta_\mu (x)), \nonumber\\ 
\tilde \Phi^{(h)}_\mu (x,
\theta, \bar\theta) &=& \phi_\mu (x) + \theta \;(s_{ab} \phi_\mu (x)) +
\bar\theta\; (s_b \phi_\mu (x)) + \theta \;\bar\theta\; (s_b s_{ab} \phi_\mu (x)),
\nonumber\\ \tilde {\cal F}^{(h)}_{\mu\nu} (x, \theta, \bar\theta) &=& C_{\mu\nu}
(x) + \theta \;(s_{ab} C_{\mu\nu} (x)) + \bar\theta\; (s_b C_{\mu\nu} (x))
+ \theta \; \bar\theta\; (s_b s_{ab} C_{\mu\nu} (x)), \nonumber\\ \tilde
{\bar {\cal F}}^{(h)}_{\mu\nu} (x, \theta, \bar\theta) &=& \bar C_{\mu\nu} (x) +
\theta\; (s_{ab} \bar C_{\mu\nu} (x)) + \bar\theta\; (s_b \bar C_{\mu\nu} (x)) + 
\theta\; \bar\theta \;(s_b s_{ab} \bar C_{\mu\nu} (x)), \nonumber\\
\tilde {\cal F}^{(h)}_1 (x, \theta, \bar\theta) &=& C_1 (x) + \theta\;(s_{ab} C_1 (x))
+ \bar \theta\; (s_b C_1 (x)) + \theta\;\bar\theta\; (s_b s_{ab} C_1 (x)), \nonumber\\
\tilde {\bar {\cal F}}^{(h)}_1 (x, \theta, \bar\theta) &=& \bar C_1 (x) + \theta\; (s_{ab}
\bar C_1 (x))
+ \bar \theta\; (s_b \bar C_1 (x)) + \theta\;\bar\theta\; (s_b s_{ab} \bar C_1 (x)), \nonumber\\
\tilde {\cal F}^{(h)}_2 (x, \theta, \bar\theta) &=& C_2 (x) + \theta\;(s_{ab} C_2 (x))
+ \bar \theta\; (s_b C_2 (x)) + \theta\;\bar\theta\; (s_b s_{ab} C_2 (x)), \nonumber\\
\tilde {\bar {\cal F}}^{(h)}_2 (x, \theta, \bar\theta) &=& \bar C_2 (x) 
+ \theta\;(s_{ab} \bar C_2 (x))
+ \bar \theta\; (s_b \bar C_2 (x)) 
+ \theta\;\bar\theta\; (s_b s_{ab} \bar C_2 (x)), 
\end{array}\eqno (A.5)
$$ where the off-shell nilpotent and absolutely anticommutating (anti-)BRST symmetry
transformations $s_{(a)b}$, with the identifications
$b_1^{(2)} = B_1, b_2^{(2)} = B_2, \bar b_2^{(1)} = \bar B,  
\bar f_\mu^{(1)} = \bar F_\mu, f_\mu^{(2)} = F_\mu,
f_\mu^{(3)} = f_\mu, \bar f_\mu^{(3)} = \bar f_\mu$, for the Abelian 3-form gauge theory are
$$
\begin{array}{lcl}
&& s_b B_{\mu\nu\eta} = \partial_\mu C_{\nu\eta} + \partial_\nu C_{\eta\mu} + \partial_\eta C_{\mu\nu},
\quad s_b C_{\mu\nu} = \partial_\mu \beta_\nu - \partial_\nu \beta_\mu, \quad s_b \bar C_{\mu\nu} = 
B_{\mu\nu}^{(2)}, \nonumber\\
&& s_b \beta_\mu = \partial_\mu C_2, \quad s_b C_2 = 0, \quad s_b B_{\mu\nu}^{(2)} = 0, \quad
s_b C_1 = - \bar B,  \quad s_b \bar B = 0, \quad s_b  B_2 = 0, \nonumber\\
&& s_b \bar C_1 = B_1, \quad s_b B_1 = 0, \quad s_b \bar C_2 = B_2, \quad s_b \bar \beta_\mu = F_\mu,
\quad s_b F_\mu = 0, \quad s_b \phi_\mu = f_\mu, \nonumber\\
&& s_b f_\mu = 0, \quad s_b \bar F_\mu = - \partial_\mu \bar B, \quad 
s_b \bar f_\mu = \partial_\mu B_1,
\quad s_b \bar B_{\mu\nu}^{(1)} = \partial_\mu f_\nu - \partial_\nu f_\mu.
\end{array}\eqno(A.6)
$$
$$
\begin{array}{lcl}
&& s_{ab} B_{\mu\nu\eta} = \partial_\mu \bar C_{\nu\eta} 
+ \partial_\nu \bar C_{\eta\mu} + \partial_\eta \bar C_{\mu\nu},
\quad s_{ab} \bar C_{\mu\nu} = \partial_\mu \bar \beta_\nu - \partial_\nu \bar\beta_\mu, 
\quad s_{ab}  C_{\mu\nu} = 
\bar B_{\mu\nu}^{(1)}, \nonumber\\
&& s_{ab} \bar \beta_\mu = \partial_\mu \bar C_2, \quad s_{ab} \bar C_2 = 0, \quad 
s_{ab} \bar B_{\mu\nu}^{(1)} = 0, \quad
s_{ab} C_1 = -  B_1,  \; s_{ab} B_1 = 0, \; s_{ab}  B_2 = 0, \nonumber\\
&& s_{ab} \bar C_1 = - B_2, \quad s_{ab} \bar B = 0, \quad s_{ab} C_2 = \bar B, 
\quad s_{ab}  \beta_\mu = \bar F_\mu,
\quad s_{ab} \bar F_\mu = 0, \; s_{ab} \phi_\mu = \bar f_\mu, \nonumber\\
&& s_{ab} \bar f_\mu = 0, \quad s_{ab} \bar F_\mu = - \partial_\mu B_2, \quad 
s_{ab}  f_\mu = - \partial_\mu B_1,
\quad s_{ab} B_{\mu\nu}^{(2)} = \partial_\mu \bar f_\nu - \partial_\nu \bar f_\mu.
\end{array}\eqno(A.7)
$$
It is elementary to check that the above 
(anti-)BRST symmetry transformations are off-shell nilpotent of order two
(i.e. $s_{(a)b}^2 = 0$).

Furthermore, it can be seen that the anticommutativity 
property (i.e. $s_b s_{ab} + s_{ab} s_b = 0$)
on the following basic  fields of the Abelian 3-form gauge theory
$$\begin{array}{lcl}
\{s_b, s_{ab} \} \;B_{\mu\nu\eta} = 0, \quad 
\{s_b, s_{ab} \} \; C_{\mu\nu} = 0, \quad
\{s_b, s_{ab} \}\; \bar C_{\mu\nu} = 0,
\end{array}\eqno(A.8)
$$
is true only when the Curci-Ferrari type restrictions (A.4) (i.e.
$ B_{\mu\nu}^{(2)} + \bar B_{\mu\nu}^{(1)} = \partial_\mu \phi_\nu - \partial_\nu \phi_\mu,
f_\mu + \bar F_\mu = \partial_\mu C_1,
\bar f_\mu + F_\mu = \partial_\mu \bar C_1 $) are satisfied. The property of the anticommutativity
of the (anti-)BRST symmetry transformations is trivially obeyed in the case of rest of the fields of 
our present 4D Abelian 3-form gauge theory.

Finally, one can write down the (anti-)BRST invariant Lagrangian density for the above
Abelian 3-form gauge theory as (see, e.g. [35,40] for details)
$$
\begin{array}{lcl}
{\cal L}^{(3)}_B &=& {\displaystyle \frac{1}{48} H^{\mu\nu\eta\xi} H_{\mu\nu\eta\xi}
+ s_b s_{ab} \Bigl ( \frac{1}{2} \bar C_1 C_1 - \bar C_2 C_2 + \frac{1}{2} \bar C_{\mu\nu} C^{\mu\nu}}
\nonumber\\
&+& \bar \beta^\mu \beta_\mu + {\displaystyle \frac{1}{2} \phi^\mu \phi_\mu - \frac{1}{6} B^{\mu\nu\eta} B_{\mu\nu\eta} \Bigr )}.
\end{array}\eqno(A.9)
$$
It will be noted that (i) all the individual terms in the big round bracket have the mass dimension two
and the ghost number equal to zero, (ii) the ghost number of the ghost fields
can be computed from the transformations (A.6) and (A.7), and (iii) the nilpotent and absolutely
anticommuating (anti-)BRST symmetry transformations (A.7) and (A.6) have been taken into account
in writing of the gauge-fixing and Faddeev-Popov ghost terms.

Our approach is quite general and can be applied to any arbitrary Abelian $p$-form gauge
theory in any arbitrary $D$-dimension of spacetime. All one has to do is generalize (A.1)
to super $p$-form connection that is defined on a (D, 2)-dimensional supermanifold. This will
automatically provide the clue about the number of basic (primary) fields of the Abelian $p$-form 
gauge theory.
The application of the celebrated horizontality condition would be able to produce the desired
nilpotent and absolutely anticommuting (anti-)BRST symmetry transformations of the theory
which can be exploited, in turn, to produce the (anti-)BRST invariant Lagrangian density
of the theory (see, e.g. (A.9)). \\

\noindent
{\bf Acknowledgement:}\\

\noindent
Financial support from the Department of Science
and Technology, Government of India, under the SERC project
grant No: SR/S2/HEP-23/2006 is gratefully acknowledged. It
is a pleasure to thank the referee for very useful suggestions.

\end{document}